\documentclass[12pt]{article} 
\hoffset = - 0.5 truecm 
 \def\lsim{\raise0.3ex\hbox{$<$\kern-0.75em\raise-1.1ex\hbox{$\sim$}}}
\def\gsim{\raise0.3ex\hbox{$>$\kern-0.75em\raise-1.1ex\hbox{$\sim$}}}
\def\beq{\begin{equation}}   \def\eeq{\end{equation}}
\def\bea{\begin{eqnarray}}  \def\eea{\end{eqnarray}} 
\usepackage{graphicx}
\usepackage{epsfig}
\begin{document}

\begin{titlepage}
 
\begin{flushright}
LAPTH-005/17\\
LPT-Orsay 16-88\\
\end{flushright}
\vspace{1.cm}

\begin{center}
\vbox to 1 truecm {}
{\large \bf  Photon-jet correlations in Deep-Inelastic Scattering}

\vskip 1 truecm
{\bf P. Aurenche$^{1,a}$, M. Fontannaz$^{2,b}$} \vskip 3 truemm

{\it $^{1}$ LAPTh, Universit\'e Savoie Mont Blanc, CNRS \\ 
BP 110, 9 Chemin de Bellevue, 74941 Annecy-le-Vieux
Cedex, France\\ 
$^{a}$ e-mail : patrick.aurenche@lapth.cnrs.fr\\
}

\vskip 3 truemm

{\it $^{2}$ Laboratoire de Physique Th\'eorique, UMR 8627 CNRS,\\
Univ. Paris-Sud, Universit\'e Paris-Saclay, 91405 Orsay Cedex, France\\ 
$^{b}$ e-mail : michel.fontannaz@th.u-psud.fr\\
}
\vskip 2 truecm

\begin{abstract}
 The reaction $e + p \to \gamma + jet + X$ is studied in QCD at the next-to-leading order. Previous studies on inclusive distributions showed a good agreement with ZEUS data. To obtain a finer understanding of the dynamics of the reaction, several correlation functions are evaluated for ZEUS kinematics.
\end{abstract}
\end{center}
\end{titlepage}

\baselineskip = 24 pt\par

This note is the continuation of the study of the Deep Inelastic Scattering (DIS) reaction $e + p \to \gamma + jet + X$ with a photon and a jet in the final state. In a preceeding paper~\cite{1r} we calculated the next-to-leading order (NLO ) QCD cross sections describing the photoproduction of a large-$p_\bot$ photon ($p_\bot^\gamma , \eta_\gamma$) accompanied by a jet. In particular, we calculated the transverse momentum and rapidity distributions of the photon with the jet constrained in a given large range as well as the jet transverse momentum spectrum in events with a detected photon. In the present note we concentrate on the distributions in the following correlation variables: $x_\gamma = (p_\bot^\gamma\ e^{- \eta_\gamma} + p_\bot ^{jet}\ e^{-\eta_{jet}})/(2 y E_e)$, $\tilde{x}_p = (p_\bot^\gamma\ e^{\eta_\gamma} + p_\bot ^{jet}\ e^{\eta_{jet}})/(2 E_p)$ (in the laboratory frame $E_e$, $E_p$ are respectively the incident energies of the electron  and the proton and $y$ is the DIS inelasticity), $\Delta \eta = \eta _\gamma - \eta_{jet}$ and $\Delta \phi = \phi_{electron} - \phi_\gamma$. Note that $\tilde{x}_p$, as defined above, does not correspond to $x_p$, the fraction of momentum carried by a parton in the proton, since we are not in a collinear frame.
The correlations should provide a finer understanding of the underlying production mechanism than the inclusive cross sections.

Originally the NLO calculation~\cite{2r,3r} was performed in the $\gamma^*-p$ center of mass ($\gamma^*$ is the virtual photon) in which a large scale is provided by a large value of $p_\bot^{*\gamma}$, the final $\gamma$ transverse momentum in that frame. Production cross sections of hadrons and jets were studied in this frame and successfully described by the NLO calculation~\cite{4r,5r,6r}. On the other hand, in the ZEUS experiment~\cite{7r} the reaction $e + p \to \gamma + jet + X$ is  studied in the laboratory frame for which the original NLO calculation must be adapted. Indeed a large $p_\bot^\gamma$ in the laboratory does not necessarily correspond to a large $p_\bot^{*\gamma}$ in the $\gamma^* - p$ frame. Therefore a cut-off $E_{\bot cut}^*$ must be introduced in the calculation with $p_\bot^{*\gamma} > E_{\bot cut}^*$ in order to remain in a perturbative domain. In the preceding paper  we considered the values  $E_{\bot cut}^*= 2.5$~GeV and $E_{\bot cut}^* = .5$~GeV. A good agreement was found between theory and the ZEUS data with the cut-off $E_{\bot cut}^* = 2.5$~GeV. We also found that there are kinematical domains in which the NLO cross sections were little dependent on $E_{\bot cut}^*$, for instance for small ratios $(Q/{{p_\bot^\gamma})^2}$ or $(Q/{{p_\bot^{jet}})^2}$. This result demonstrates that the theoretical calculations are almost model independent in the corresponding domains. A detailed discussion of this problem can be found in ref.~\cite{1r}. 

In this note we pursue the study of the reaction $e+p \to \gamma +jet + X$ by calculating the cross sections $d\sigma /dx_\gamma$, $d\sigma /dx_{\tilde{x}_p}$, $d\sigma /d\Delta \eta$ and $d\sigma /d\Delta \phi$ with the cut-offs $E_{\bot cut}^* = 2.5$~GeV and $E_{\bot cut}^* = .5$~GeV, and, following ZEUS~\cite{7rPB}, we consider two $Q^2$-ranges, namely $10 < Q^2 < 350$ GeV$^2$ and $10 < Q^2 < 30$~GeV$^2$. As in ref.~\cite{1r} we adopt the ZEUS kinematics~\cite{7r} with $\sqrt{s} = 319$~GeV; the photon momentum has to lie in the ranges $4 < p_\bot^\gamma < 15$~GeV and $-.7 < \eta_\gamma < .9$. For the jet momentum we have $2.5 < p_\bot^{jet} < 35$~GeV and $-1.5 < \eta_{jet} < 1.8$. Constraints on the final electron are~: $E{_e'} > 10$~GeV, $\theta_e > 140^\circ$ (the $z$ axis is pointing toward the proton direction). As in the ZEUS experiment, the photon is isolated using the democratic $k_\perp$-algorithm~\cite{GloverMorgan} where the photon is treated on the same footing as partons. We refer to ref.~\cite{1r} for details on the isolation criteria and other parameters and conventions used in the calculation.  We do not consider here the emission of photons by the electron. Therefore, this cross section should be added to those calculated in this paper to reconstruct the experimental cross section measured by the ZEUS collaboration.\\

We recall that the cross section is the sum of four building blocks. The "direct" component where the initial virtual photon is coupled to the hard partonic process and the "resolved" component where it interacts via its structure fonction. In each of these the final real photon couples directly to the hard process or is a fragment of a parton produced at large transverse momentum (fragmentation component). Due to the photon isolation criteria the fragmentation components are small, typically 10\% to 20\% of the total depending on the kinematics. We use the virtual photon structure function presented in ref.~\cite{9r} and the CTEQ6M parton distributions in the proton~\cite{8r}. The fragmentation function is that of the BFG collaboration (set II)~\cite{10r}. All the components are calculated at NLO and details can be found in ref.~\cite{11r}. We work in the $\overline{\rm MS}$ scheme for factorisation and renormalisation with $\Lambda_{\overline{MS}}(4) = 236$~MeV and $N_f = 4$. All scales are taken equal to $\sqrt{{p_{\bot}^{*\gamma}}^{2} + Q^2}$. The numerical calculations are carried out using the adaptive Monte Carlo code BASIS~\cite{12r}.

The results of the calculations are displayed in Fig.~1 to Fig.~4. Each figure contains four curves corresponding to two different values of $E_{\bot cut}^*$ and to two $Q^2$-integration domains. 
\begin{figure}
\vspace{9pt}
\centering
\includegraphics[scale=0.8]{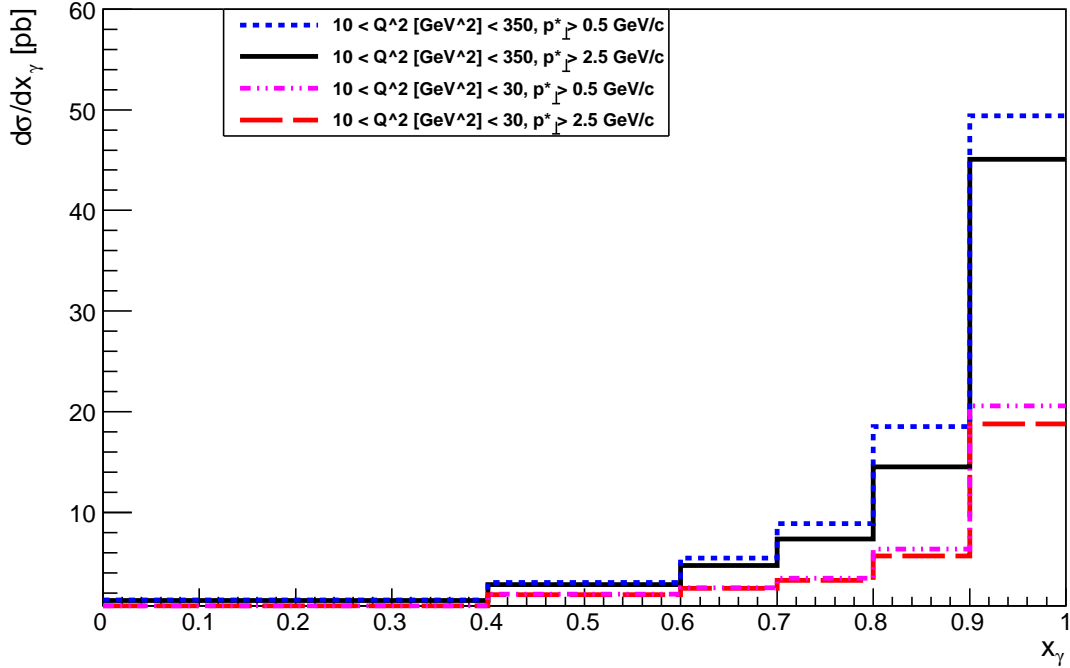}
\caption{$d\sigma /dx_\gamma$ cross sections for the ZEUS kinematics~\cite{7r}.} 
\label{fig:1}
\end{figure}
\begin{figure}[t]
\centering
\includegraphics[scale=0.8]{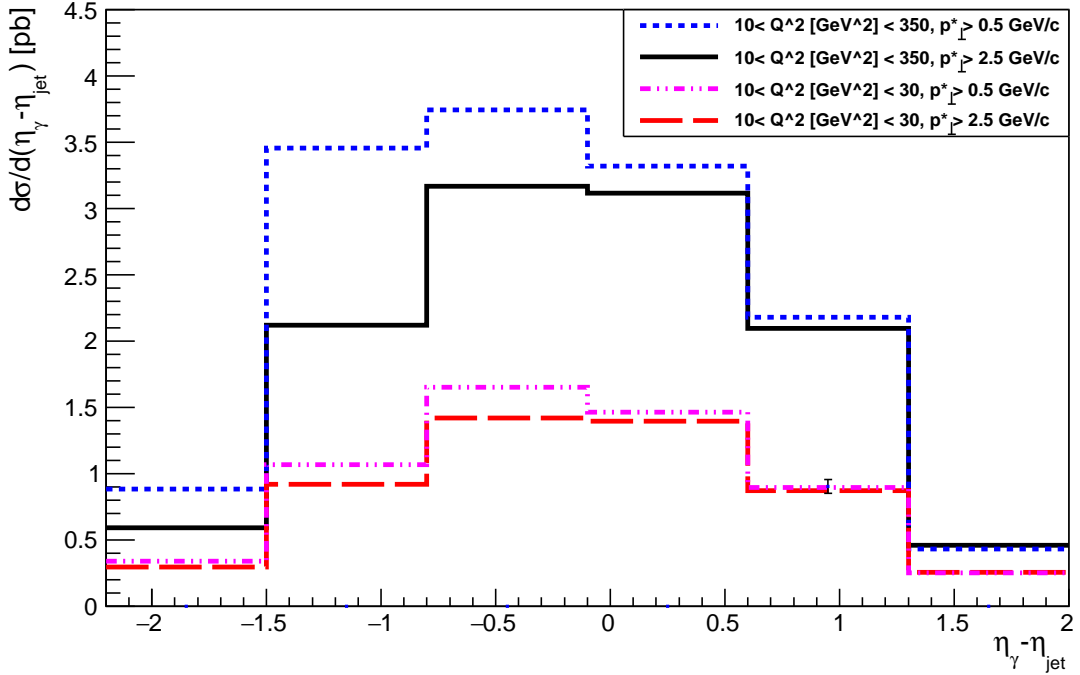}
\caption{$d\sigma /d\Delta \eta$ cross sections for the ZEUS kinematics~\cite{7r}. The point at $\eta_\gamma~-~\eta_{jet}~=~.95$ with an error bar is an indication of the variation of the theoretical prediction when we vary the coefficient in the scales in the range $.5 < c < 2.$ (see the text).} 
\label{fig:2}
\end{figure}
The shapes of the curves corresponding to the small domain $10 < Q^2< 30$ GeV$^2$ (red curves) and to $10 < Q^2  < 350$ GeV$^2$ are similar, but the curves of the small $Q^2$-domain are a factor 2 to 3 smaller than those of the full $Q^2$ range. For each $Q^2$ domain two values of $E_{\bot cut}^*$ have been used, $E_{\bot cut}^*=.5$~GeV and $E_{\bot cut}^*=2.5$~GeV. It clearly appears that the low $Q^2$ data are globally less sensitive to the transverse momentum cut-off $E_{\bot cut}^*$~: indeed, in the laboratory, the transverse momentum of the virtual photon is $\sqrt{Q^2 (1-y)}$ and, therefore, at small $Q^2$ the values of $p_\bot^{\gamma}$ and $p_\bot^{*\gamma}$ are closer and the laboratory constraint $p_\bot^\gamma > 4.$~GeV is more efficient in cutting off the small $p_\bot^{*\gamma}$ configurations. 
\begin{figure}
\vspace{9pt}
\centering
\includegraphics[scale=0.8]{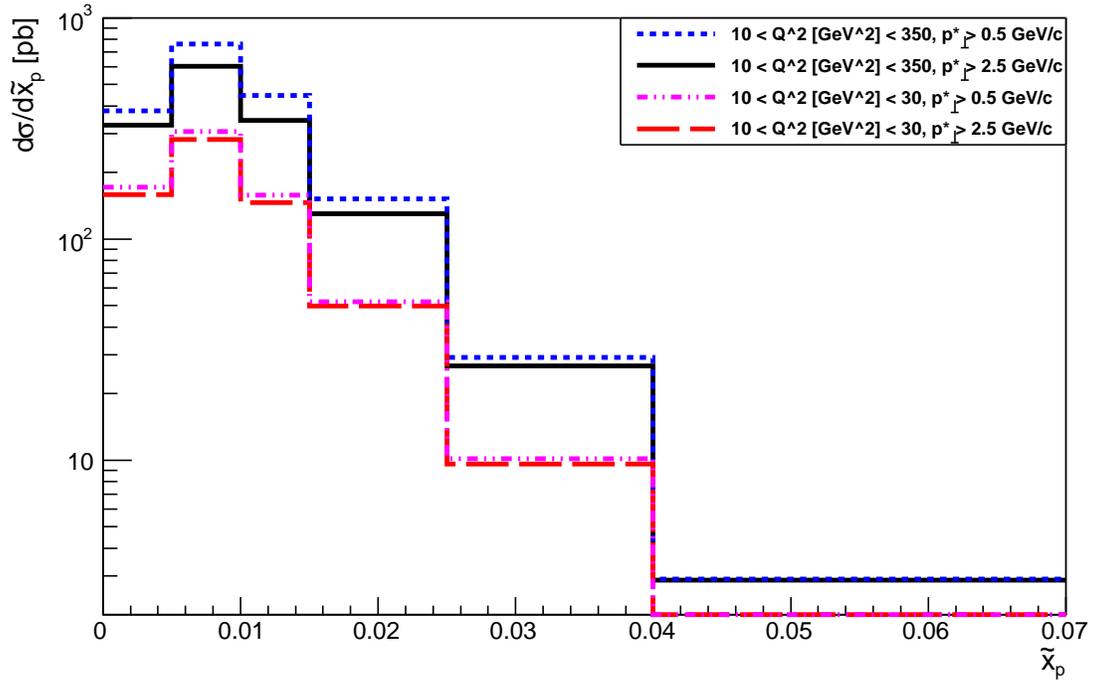}
\caption{$d\sigma /d\tilde{x}_p$ cross sections for the ZEUS kinematics~\cite{7r}.} 
\label{fig:3}
\end{figure}
\begin{figure}
\vspace{9pt}
\centering
\includegraphics[scale=0.8]{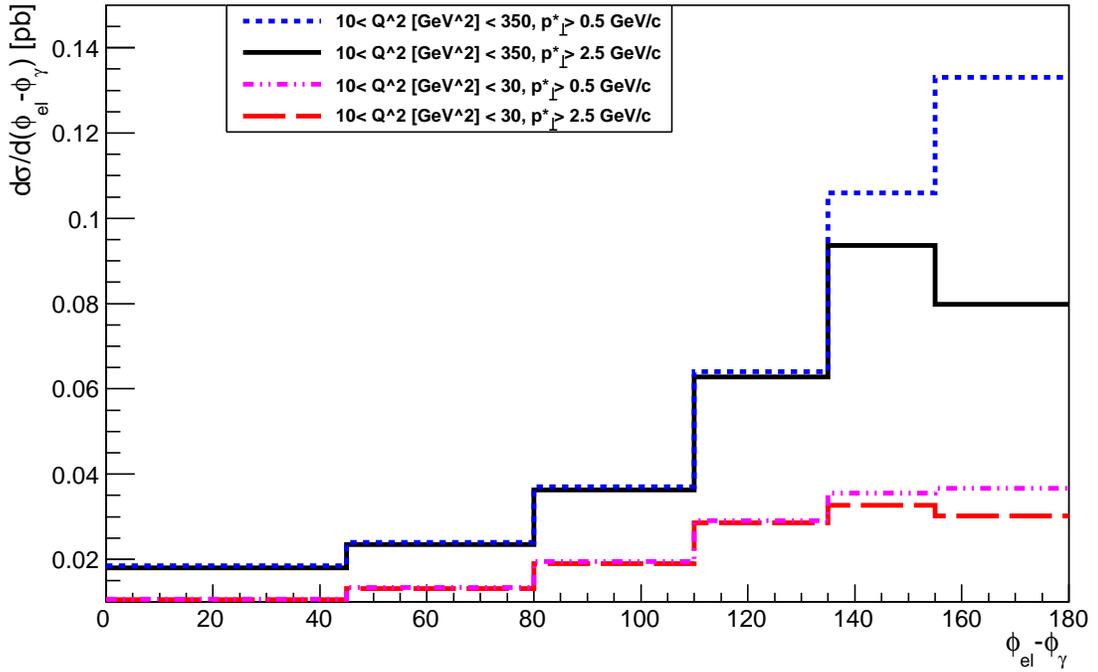}
\caption{$d\sigma /d\Delta \phi$ cross sections for the ZEUS kinematics~\cite{7r}.} 
\label{fig:4}
\end{figure}
For the large $Q^2$ case the effect of the $E_{\bot cut}^*$-variation is irregular and strongly depends on the values of the kinematical variables. Therefore the shapes of the distributions are sensitive to $E_{\bot cut}^*$. But there are kinematical domains in which the distributions are little affected by $E_{\bot cut}^*$. For instance, $d\sigma /dx_\gamma$ is rather independent on $E_{\bot cut}^*$ for $x_\gamma \ \lsim \ .7$. This can be understood in the following way~: a small value of $x_\gamma$ corresponds to a large value of $y$ and consequently a small transverse momentum of the initial virtual photon. Also the parton with a small $x_\gamma$\footnote{The Born direct contribution corresponds to $x_\gamma = 1.$} carries away a small transverse momentum. These two effects explain the weak sensitivity of $d\sigma /dx_\gamma$ to $E_{\bot cut}^*$ for $x_\gamma \ \lsim \ .7$. Likewise, the positive $\Delta \eta = \eta _\gamma - \eta_{jet}$ domain is rather stable : 
in this domain the cross section is dominated by the direct processes, less sensitive to $E_{\bot cut}^*$ than the resolved contributions.  As for $d\sigma /d\tilde{x}_p$ a similar reason leads to a stability region for large $\tilde{x}_p$. A stable behavior is observable for $d\sigma /d\Delta \phi$ in the domain $\Delta \phi < 140^\circ$, which can be explained by kinematical reasons. At $\Delta \phi = 0$, for instance, $p_\perp^{*\gamma}$ is always larger than $p_\bot^{\gamma} > 4$~GeV and an $E_{\bot cut}^*$ is not necessary. When there is a large sensitivity, we found in ref.~\cite{1r}  that inclusive data were well described with the choice $E_{\bot cut}^* = 2.5$~GeV.

Concerning the sensitivity of the cross-sections to the factorisation and renormalisation scales, we have studied it for one representative point of the distributions in the case of $10 < Q^2 < 30$ GeV$^2$. For the renormalisation and the proton factorisation we use 
$\mu = M = c \sqrt{Q^2 + {p_\bot^{*\gamma}}^2}$ while in the virtual photon structure function we take $M_F = \sqrt{Q^2 + (c {p_\bot^{*\gamma}})^2}$. This last choice is natural since $M_F$ cannot be smaller than $\sqrt{Q^2}$ in the photon structure function~\cite{9r}. The sample points are $.6 < x_\gamma < .7$, $.6 < \Delta\eta < 1.3 $ and $80^\circ < \Delta\phi < 110^\circ$, domains in which the cross sections are almost insensitive to $E_{\bot cut}^*$. We observe that in the range $.5 < c < 2.$ the cross section varies by at most 15\%. For example, for the distribution in $x_\gamma$ this variation corresponds to the thickness of the line in Fig. 1, while for the distribution in $\eta_\gamma - \eta_{jet}$ it is displayed on Fig. 2, at the value $\eta_\gamma - \eta_{jet} = .95$.
We notice that this relative stability is the result of a huge compensation between the direct and the resolved terms whose size varies by a factor 2 to 4 under the scales changes.\

Finally, concerning the accuracy of the theoretical calculations, let us discuss another point besides the sensitivity to the cut $E_{\bot cut}^*$ and to the scale variations. When calculating the higher order corrections to the resolved component, we assume that the photon structure function, proportional to $\ln {p_\bot^{*2} + Q^2 \over Q^2}$, is large, so that we can neglect terms not proportional to this logarithm. We have no way to estimate the importance of such terms in the ZEUS experiment in which the condition
$Q^2/p_\bot^{*2} \ll 1$ is not always verified. To test the validity of the approximation it would be interesting to compare theory and data in the two ranges $10<Q^2<30$ GeV$^2$ and $30<Q^2<350$ GeV$^2$~\cite{13r}.

In conclusion, in the ZEUS experimental configuration for observables constructed with the laboratory kinematics we have a good theoretical stability under changes of cut-offs in a relatively large domain  of the variables $x_\gamma< 0.7$, $\Delta\eta > 0$ and $\Delta\phi < 130^\circ$  when $10 < Q^2  < 350$ GeV$^2$. Outside these ranges the physics becomes sensitive to non perturbative effects. In the small $Q^2$ domain the QCD predictions are independent on the cut-offs in the above quoted ranges and rather insensitive elsewhere. As for the scale uncertainties they are less than $\pm 8 \%$ when probing the standard scale ranges. Comparison of these NLO predictions with preliminary ZEUS data are in good agreement after the photon emission from the electron is taken into account~\cite{13r}.\\

\noindent{\bf Acknowledgments}\\
We thank Peter Bussey for useful discussions.

\end{document}